\documentclass[12pt]{article}

\usepackage[numbers,sort&compress]{natbib}
\usepackage{amsmath}
\usepackage{amssymb}
\usepackage{hyperref}
\usepackage[bf]{caption}
\usepackage{color}

\usepackage{sectsty}
\sectionfont{\large}
\subsectionfont{\normalsize}

\usepackage{float}

\newcommand{\ba}{\begin{aligned}}
\newcommand{\ea}{\end{aligned}}
\newcommand{\beq}{\begin{equation}}
\newcommand{\eeq}{\end{equation}}
\def\ben{\begin{equation*}}
\def\een{\end{equation*}}
\newcommand{\beqs}{\begin{eqnarray}}
\newcommand{\eeqs}{\end{eqnarray}}

\newcommand{\sdot}{\hspace{-3pt}\cdot\hspace{-3pt}}

\def\be{\begin{equation}}
\def\ee{\end{equation}}
\def\bea{\begin{eqnarray}}
\def\eea{\end{eqnarray}}
\def\bsp{\be\begin{split}}

\newcommand{\underbracedmatrix}[2]{%
  \left(\;
  \smash[b]{\underbrace{
    \begin{matrix}#1\end{matrix}
  }_{#2}}
  \;\right)
  \vphantom{\underbrace{\begin{matrix}#1\end{matrix}}_{#2}}
}

\def\a{\alpha}
\def\b{\beta}

\def\d{\delta}
\def\e{\epsilon}

\def\s{\sigma}

\makeatletter

\newcommand{\Rmnum}[1]{\expandafter\@slowromancap\romannumeral #1@}
\makeatother

\renewcommand{\title}[1]{\vbox{\center\LARGE{#1}}\vspace{5mm}}
\renewcommand{\author}[1]{\vbox{\center\large{#1}}\vspace{5mm}}

\everymath{\displaystyle}

\begin{document}

\begin{titlepage}
\begin{flushright}
\vspace{10pt} \hfill{NCTS-TH/1506} \vspace{20mm}
\end{flushright}
\begin{center}

{\Large \bf Elimination and recursions in the scattering equations}

\vspace{45pt}

{
\textbf{Carlos Cardona},$^a$ 
\textbf{Chrysostomos Kalousios},$^b$
\footnote[1]{
\href{mailto:carlosandres@mx.nthu.edu.tw }{\tt{carlosandres@mx.nthu.edu.tw}}\,,  
\href{mailto:ckalousi@ift.unesp.br}{\tt{ckalousi@ift.unesp.br}}
}
}
\\[15mm]

{\it\ 
${}^a\,$Physics Division, National Center for Theoretical Sciences, National Tsing-Hua University,
Hsinchu, Taiwan 30013, Republic of China.

\vspace{20pt}

${}^b\,$ICTP South American Institute for Fundamental Research\\
Instituto de F\'\i sica Te\'orica, UNESP-Universidade Estadual Paulista\\
R. Dr. Bento T. Ferraz 271 - Bl. II, 01140-070, S\~ao Paulo, SP, Brasil}\\
\vspace{10pt}

\vspace{20pt}

\end{center}

\vspace{40pt}

\centerline{{\bf{Abstract}}}
\vspace*{5mm}
\noindent
We use the elimination theory to explicitly construct the $(n-3)!$ order polynomial in one of the variables of the scattering equations.  The answer can be given either in terms of a determinant of Sylvester type of dimension $(n-3)!$ or a determinant of B\'{e}zout type of dimension $(n-4)!$.  We present a recursive formula for the Sylvester determinant.  Expansion of the determinants yields expressions in terms of Pl\"{u}cker coordinates.  Elimination of the rest of the variables of the scattering equations is also presented.

\vspace{15pt}
\end{titlepage}

\newpage   

\section{Introduction}
The last decade a great amount of progress has been made in our understanding of scattering amplitudes for massless theories starting with the revolutionary work of Witten \cite{Witten:2003nn} on four dimensional scattering amplitudes in twistor space. In 2013 a new formulation for the computation of the tree level $S$-matrix of massless particles in any dimension was proposed in a series of papers by Cachazo, He and Yuan (CHY) \cite{Cachazo:2013gna,Cachazo:2013hca,Cachazo:2013iea,Cachazo:2014nsa,Cachazo:2014xea}. The formalism was proven for Yang-Mills and scalars in \cite{Dolan:2013isa}.  By defining a map from the space of kinematic invariants of the scattering of $n$ massless particles to the space of punctures over a Riemann sphere, CHY found the so-called scattering equations,
\be\label{scat eq}
f_a \equiv \sum_{b\neq a}^n\frac{k_a\sdot k_b}{\s_a-\s_b}=0,\quad a=1,2,\ldots,n, 
\ee  
where $k_a$ denotes the momentum of the $a^{\rm th}$ particle and $\s_b$ the position of the $b^{\rm th}$ puncture on the sphere. 
Those equations have appeared previously in different contexts, particularly in the work of Fairlie and Roberts \cite{Fairlie,Roberts,Fairlie:2008dg} and in the study of the high energy behavior of scattering of strings by Gross and Mende \cite{Gross:1987ar} (see also  \cite{Witten:2004cp, Caputa:2011zk, Caputa:2012pi, Makeenko:2011dm, Cachazo:2012uq}).
In this work we use the polynomial form of \eqref{scat eq} that was presented in \cite{Dolan:2014ega} and which can be written as
\be\label{fixed gauge scattering equations} 
\sum_{S \subset A,~|S|=m} k_S^2 \s_S = 0, \qquad 2 \leq m \leq n-2,
\ee
where $A={1,2,\ldots,n}$ and the sum runs over all subsets of $S$ with $m$ elements.  Furthermore,
\be  
k_S  = \sum_{a\in S} k_a, \qquad \s_S = \prod_{b\in S}\s_b.
\ee
According to CHY, massless scattering at tree level in arbitrary dimensions can in general be described by the contour integral
\be \label{An=}
\mathcal{A}_n = \int \frac{{\rm d}^n \s}{{\rm vol}\, SL(2,\mathbb{C})}
\s_{ij}\s_{jk}\s_{ki}\hspace{-6pt} \prod_{a \neq i,j,k}\hspace{-6pt} \d(f_a) \hspace{2pt} I_n(k,\e,\s),
\ee
which is completely localized over the solutions of the scattering equations, and  $I_n(k,\e,\s)$ depends on the theory. So far $I_n(k,\e,\s)$ has been proposed for several different theories in various dimensions, including Einstein gravity, pure Yang-Mills \cite{Cachazo:2013hca}, $\phi^3$ \cite{Cachazo:2013iea}, Einstein-Yang-Mills \cite{Cachazo:2014nsa}, massive scalar-gravity \cite{Naculich:2014naa} and lately for Dirac-Born-Infeld and $U(N)$ non-linear sigma model \cite{Cachazo:2014xea}. 

Since the CHY original formulation, many papers have surfaced  investigating several aspects of the scattering equations, such as its relation to momentum twistor formalism \cite{Weinzierl:2014vwa}, solutions at particular kinematics (for example \cite{Cachazo:2013iea, Dolan:2014ega, Kalousios:2013eca, Weinzierl:2014vwa, Lam:2014tga}), its geometrical interpretation \cite{He:2014wua}, soft limits \cite{Schwab:2014xua, Cachazo:2015ksa,Afkhami-Jeddi:2014fia, Kalousios:2014uva}, generalizations to include massive particles (for example \cite{Naculich:2014naa,Naculich:2015zha,Naculich:2015coa,
Weinzierl:2014ava, delaCruz:2015raa}) and formulations in terms of string-like models (for example \cite{Mason:2013sva,Berkovits:2013xba,Bjerrum-Bohr:2014qwa,Casali:2015vta}).

Although at first sight the CHY formalism seems quite simple and straightforward, when it comes to the study of solutions of the scattering equations they rapidly become difficult to manage since the number of solutions grows factorially as a function of the number of particles. However the amplitudes involve a sum over the solutions which can be found without even solving the equations.
With that in mind, several interesting methods have been developed lately. In \cite{Kalousios:2015fya} the elimination method was used in order to write the scattering amplitude in terms of the solutions of a polynomial in a single variable, $\s_a$, and then use the Vieta formulas to get the sums over roots from the coefficients of the one variable polynomial in $\s_a$. This idea was enhanced in \cite{Huang:2015yka} by using the companion matrix method, which allows to encode the single variable polynomial in a matrix $T_{\s_a}$ such that its eigenvalues correspond to the roots of the polynomial and hence the sum over the roots is reduced to taking the trace of the given matrix. In \cite{Cardona:2015eba} we further elaborated on the equivalence of \cite{Kalousios:2015fya} and \cite{Huang:2015yka}. A related approach has been recently considered in \cite{Sogaard:2015dba} where the authors use the B\'ezoutian matrix to compute the amplitudes with emphasis to numerics. A parallel method was developed in \cite{Cachazo:2015nwa} where known results for $\phi^3$ theory are used as building blocks for the computation of other amplitudes. A different approach was considered in \cite{Baadsgaard:2015voa,Baadsgaard:2015ifa}, where based on some results from \cite{Bjerrum-Bohr:2014qwa} a direct matching between Feynman diagrams and integration measures in the scattering equation formalism of CHY was found. 

This work is partly motivated by some ideas discussed in \cite{Kalousios:2015fya} and then in \cite{Huang:2015yka} and \cite{Cardona:2015eba}.  If one can eliminate the variables in the scattering equations, then using simple algebra one can evaluate any rational expression of the variables of the scattering equations summed over roots.  Of most importance are of course the scattering amplitudes.  The elimination theory was first used in the $n=5$ and $n=6$ case in the work of \cite{Dolan:2014ega}.  Here we are concerned with the general case.

We know from B\'{e}zout theorem that the number of solutions of \eqref{fixed gauge scattering equations} is $(n-3)!$.  This means that in the general case each of the variables appearing in the scattering equations can be expressed as one variable polynomials of order $(n-3)!$.  In the following we will show how this can be done for the chosen variable $\s_{n-1}$.  We will then express the rest of the variables of the scattering equations as functions of $\s_{n-1}$ only.  In the mathematical literature this goes by the name of elimination theory.  We will study the elimination theory of \eqref{fixed gauge scattering equations} following two separate approaches, by Sylvester and B\'{e}zout.  The results obtained are the same and can be expressed in various determinantal forms.  On our way we will also discover an on-shell recursion for the scattering equations.

In the whole body of this paper we think of \eqref{fixed gauge scattering equations} as a special case of the multilinear system of equations
\be \label{generalized scattering equations 1}
g''_m \equiv \sum_{i_j \in \{0,1\} }b_{i_3 i_4 \ldots i_{n-1} m}\s_3^{i_3} \s_4^{i_4} \cdots \s_{n-1}^{i_{n-1}} = 0, \qquad m=1,2,\ldots n-3,
\ee
where $b$ are arbitrary constants and we have fixed the $SL(2,\mathbb{C})$ invariance of \eqref{scat eq} by specifying arbitrary values to $\s_1,\s_2,\s_n$.  It is notationally convenient for what follows to consider $\s_{n-1}$ as a parameter and rewrite \eqref{generalized scattering equations 1} as
\be \label{generalized scattering equations 2}
g_m \equiv \sum_{i_j \in \{0,1\}} c_{i_3 i_4 \ldots i_{n-2} m}\s_3^{i_3} \s_4^{i_4} \cdots \s_{n-2}^{i_{n-2}} = 0, \qquad m=1,2,\ldots n-3,
\ee
where $c$ now depends linearly on $\s_{n-1}$.  This defines a system of $n-3$ equations in $n-4$ unknowns.  We now turn to the elimination via Sylvester type determinants.

\section{Elimination in the scattering equations via Sylvester type determinants}
In this part we will express the advertised relations in terms of Sylvester type determinants.  Before focusing on the general case, we will first discuss the $n=6$, $n=7$ and $n=8$ case.  The  $n=5$ case is trivial.

\subsection{$n=6$}
The number of scattering equations is three in this case.  Let $S(a_1,a_2,\ldots, a_i)$ denote the vector space of all polynomials of degree less than or equal to $a_i$ in $x_i$.  Then the dimension of $S(a_1,a_2,\ldots, a_i)$ is $\prod_{j=1}^i(a_j+1)$.  We consider the linear map
\be 
\phi_3 ~ : ~ S(0,1)^3 \rightarrow S(1,2),~ (f_1,f_2,f_3) \mapsto f_1 g_1 + f_2 g_2 + f_3 g_3,
\ee
where $g_i$ are given by \eqref{generalized scattering equations 2} for $n=6$.  Both the range and image of $\phi_3$ have dimension 6.  We choose to use the canonical order for the monomial bases for both vector spaces.  We choose $\s_3$ and $\s_4$ to be the variables of the scattering equations and we consider $\s_6$ to be a parameter.  Then $\phi_3$ is given by a $6\times6$ matrix whose determinant is the resultant of the three scattering equations.  By setting the resultant to zero we get the desired sixth order polynomial in $\s_5$.  We proceed with the explicit construction of the resultant.

From \eqref{generalized scattering equations 2} the three scattering equations are
\be\label{n6 scattering eqs} 
g_i = c_{00i}+c_{10i}\s_3 + c_{01i} \s_4 + c_{11i} \s_3 \s_4 = 0, \qquad i=1,2,3.
\ee
The monomial basis of $S(0,1)$ is $\{1\} \otimes \{1, \s_4 \} = \{1,\s_4\} $ and that of $S(1,2)$ is  $\{1,\s_3\} \otimes \{1, \s_4, \s_4^2 \} = \{1,\s_3,\s_4 , \s_3\s_4, \s_4^2, \s_3 \s_4^2\} $.  We multiply each of the three scattering equations with the elements of the monomial basis $S(0,1)$ and we consider the following system of six equations 
\be \label{n=3 6 scattering equations}
g_1 = g_2 = g_3 = \s_4 g_1 = \s_4 g_2 = \s_4 g_3 = 0.
\ee
Then \eqref{n=3 6 scattering equations} is a linear system in the variables of the monomial basis $S(1,2)$.  The desired sixth order polynomial in $\s_5$ is then given by the determinant of \eqref{n=3 6 scattering equations} and it is
\be\label{6x6 determinant}
\begin{vmatrix}
c_{001} & c_{002} & c_{003} & 0 & 0 & 0 \\
c_{101} & c_{102} & c_{103} & 0 & 0 & 0 \\
c_{011} & c_{012} & c_{013} & c_{001} & c_{002} & c_{003} \\
c_{111} & c_{112} & c_{113} & c_{101} & c_{102} & c_{103} \\
0 & 0 & 0 & c_{011} & c_{012} & c_{013}\\
0 & 0 & 0 & c_{111} & c_{112} & c_{113}
\end{vmatrix}
=
\begin{vmatrix}
g_1 & g_2 & g_3 & 0 & 0 & 0 \\
g_{1,3} & g_{2,3} & g_{3,3} & 0 & 0 & 0 \\
g_{1,4} & g_{2,4} & g_{3,4} & g_1 & g_2 & g_3 \\
g_{1,34} & g_{2,34} & g_{3,34} & g_{1,3} & g_{2,3} & g_{3,3} \\
0 & 0 & 0 & g_{1,4} & g_{2,4} & g_{3,4} \\
0 & 0 & 0 & g_{1,34} & g_{2,34} & g_{3,34} 
\end{vmatrix}
=0,
\ee
where $g_{i,i_1 i_2 \cdots i_r} = \frac{\partial^r g_i}{\partial \s_{i_1} \partial \s_{i_2} \ldots \partial \s_{i_r}}$.  We can also rewrite \eqref{6x6 determinant} compactly as
\be 
\begin{vmatrix}
A_0 & 0 \\
A_1 & A_0 \\
0 & A_1
\end{vmatrix}
=0, \qquad
A_{j_1} = 
\begin{pmatrix}
c_{0j_1 1} & c_{0j_1 2} & c_{0j_1 3} \\
c_{1j_1 1} & c_{1j_1 2} & c_{1j_1 3}
\end{pmatrix}, \qquad j_1=0,1.
\ee

There are many ways to expand \eqref{6x6 determinant}, for example we can consider the Laplace expansion of the first three columns to get
\be 
\eqref{6x6 determinant} = \left[ 00,10,01 \right] \left[ 10, 01, 11 \right] -
\left[00,10,11\right] \left[00,01,11\right] =
\begin{vmatrix}
\left[ 00,10,01 \right] & \left[00,10,11\right] \\
\left[00,01,11\right] &  \left[ 10, 01, 11 \right]
\end{vmatrix} ,
\ee
where we have used the $n$-bracket notation
\be\ba\label{n-bracket}
&\left[ i_{11} i_{12}\ldots i_{1(n-1)}, i_{21} i_{22}\ldots i_{2(n-1)} , \ldots, i_{n1} i_{n2}\ldots i_{n(n-1)} \right] =\\[0.2cm]
& \qquad \qquad 
\begin{vmatrix}
c_{i_{11} i_{12}\ldots i_{1(n-1)} 1} & c_{i_{11} i_{12}\ldots i_{1(n-1)} 2} & \cdots & c_{i_{11} i_{12}\ldots i_{1(n-1)} n} \\
c_{i_{21} i_{22}\ldots i_{2(n-1)} 1} & c_{i_{21} i_{22}\ldots i_{2(n-1)} 2} & \cdots & c_{i_{21} i_{22}\ldots i_{2(n-1)} n} \\
\vdots & \vdots & \ddots & \vdots \\
c_{i_{n1} i_{n2}\ldots i_{n(n-1)} 1} & c_{n_{11} i_{n2}\ldots i_{n(n-1)} 2} & \cdots & c_{i_{n1} i_{n2}\ldots i_{n(n-1)} n}
\end{vmatrix}.
\ea\ee

In order to complete the elimination we need to express $\s_3$ and $\s_4$ as function of $\s_5$.  This can be achieved in various ways.  We can consider any five of the six equations in \eqref{n=3 6 scattering equations}, for example the first five, and use Cramer's rule to solve the corresponding linear system for $\s_3$ and $\s_4$.  We respectively find
\be\ba\label{5x5 determinant}
&\begin{vmatrix}
c_{001} +  c_{101} \s_3 & c_{002}+ c_{102} \s_3 & c_{003} +  c_{103} \s_3& 0 & 0  \\
c_{011} & c_{012} & c_{013} & c_{001} & c_{002}  \\
c_{111} & c_{112} & c_{113} & c_{101} & c_{102}  \\
0 & 0 & 0 & c_{011} & c_{012} \\
0 & 0 & 0 & c_{111} & c_{112} 
\end{vmatrix} = \\[0.2cm]
&
\begin{vmatrix}
c_{001} +  c_{011} \s_4 & c_{002}+ c_{012} \s_4 & c_{003} +  c_{013} \s_4 &  c_{001} \s_4 &  c_{002} \s_4 \\
c_{101} & c_{102} & c_{103} & 0 & 0  \\
c_{111} & c_{112} & c_{113} & c_{101} & c_{102}  \\
0 & 0 & 0 & c_{011} & c_{012} \\
0 & 0 & 0 & c_{111} & c_{112} 
\end{vmatrix}
=0
\ea\ee
The two determinants in \eqref{5x5 determinant} came from \eqref{6x6 determinant} by removing the sixth column which corresponds to the sixth scattering equation in \eqref{n=3 6 scattering equations} and by multiplying the second (third) row by $\s_3$ ($\s_4$) and adding that to the first row, finally eliminating the second (third) row.

In the case of $n=6$ it is easy to simplify the problem even further by noticing that the first three equations in \eqref{n=3 6 scattering equations} only depend on three variables ($\s_3,\s_4,\s_3\s_4$).  Therefore, it is sufficient to only consider the top left $4\times 3$ block of \eqref{6x6 determinant} and apply the same operations that led to \eqref{5x5 determinant}.  Equivalently we can only consider the top left $3\times 3$ blocks of \eqref{5x5 determinant}. More specifically we find
\be \ba
& \begin{vmatrix}
c_{001} + c_{101} \s_3 & c_{002} + c_{102} \s_3 & c_{003} + c_{103} \s_3 \\
c_{011} & c_{012} & c_{013} \\
c_{111} & c_{112} & c_{113}
\end{vmatrix} =
\begin{vmatrix}
g_1 & g_2 & g_3 \\
g_{1,4} & g_{2,4} & g_{3,4} \\
g_{1,34} & g_{2,34} & g_{3,34}
\end{vmatrix} = 0, \\[0.2cm]
& \begin{vmatrix}
c_{001} + c_{011} \s_4 & c_{002} + c_{012} \s_4 & c_{003} + c_{013} \s_4 \\
c_{101} & c_{102} & c_{103} \\
c_{111} & c_{112} & c_{113}
\end{vmatrix} =
\begin{vmatrix}
g_1 & g_2 & g_3 \\
g_{1,3} & g_{2,3} & g_{3,3} \\
g_{1,34} & g_{2,34} & g_{3,34}
\end{vmatrix} = 0.
\ea \ee

\subsection{$n=7$}
We consider the map
\be 
\phi_4 ~ : ~ S(0,1,2)^4 \rightarrow S(1,2,3),~ (f_1,f_2,f_3,f_4) \mapsto f_1 g_1 + f_2 g_2 + f_3 g_3 + f_4 g_4.
\ee
In this case the range and image of $\phi_4$ have dimension 24.  The monomial basis of $S(0,1,2)$ is $\{1\} \otimes \{1, \s_4 \} \otimes \{1, \s_5 , \s_5^2 \} = \{1,\s_4, \s_5, \s_4 \s_5, \s_5^2, \s_4 \s_5^2 \} $ and that of $S(1,2,3)$ is  $\{1,\s_3\} \otimes \{1, \s_4, \s_4^2 \} \otimes \{1, \s_5, \s_5^2, \s_5^3 \} = \{1,\s_3,\s_4 , \s_3\s_4, \s_4^2, \s_3 \s_4^2 , \s_5, \s_3 \s_5, \s_4 \s_5, \s_3 \s_4 \s_5, \s_4^2 \s_5, \s_3 \s_4^2 \s_5, \\ \s_5^2, \s_3  \s_5^2, \s_4 \s_5^2, \s_3 \s_4 \s_5^2, \s_4^2 \s_5^2, \s_3 \s_4^2 \s_5^2, \s_5^3, \s_3 \s_5^3, \s_4 \s_5^3, \s_3 \s_4 \s_5^3, \s_4^2 \s_5^3, \s_3 \s_4^2 \s_5^3 \} $.

From \eqref{generalized scattering equations 2} the four scattering equations are
\be\ba\label{n7 scattering eqs}
g_i =  c_{000i}&+c_{100i}\s_3 + c_{010i} \s_4 + c_{001i} \s_5 + c_{110i} \s_3 \s_4 \\
&+ c_{101i} \s_3 \s_5 + c_{011i} \s_4 \s_5 + c_{111i} \s_3 \s_4 \s_5 = 0,\qquad i=1,2,3,4.
\ea\ee
We consider the following set of 24 equations
\be 
g_i = \s_4 g_i = \s_5 g_i = \s_4 \s_5 g_i = \s_5^2 g_i = \s_4 \s_5^2 g_i = 0, \qquad i=1,2,3,4
\ee
and we expand them in the monomial basis of $S(1,2,3)$.  Then the sought 24 degree polynomial in $\s_6$ is given by the $24\times 24$ determinant
\be\label{24x24 Sylvester}
\begin{vmatrix}
B_0 & 0 & 0 \\
B_1 & B_0 & 0 \\
0 & B_1 & B_0 \\
0 & 0 & B_1
\end{vmatrix}
=0,
~
B_{j_1}=
\begin{pmatrix}
A_{0j_1} & 0 \\
A_{1j_1} & A_{0j_1} \\
0 & A_{1j_1}
\end{pmatrix},
~
A_{j_1 j_2} = 
\begin{pmatrix}
c_{0j_1 j_2 1} & c_{0j_1 j_2 2} & c_{0j_1 j_2 3} & c_{0j_1 j_2 4} \\
c_{1j_1 j_2 1} & c_{1j_1 j_2 2} & c_{1j_1 j_2 3} & c_{1j_1 j_2 4}
\end{pmatrix}.
\ee
The range of $j_i$ is 0,1.

We now need to express $\s_3,\s_4,\s_5$ as functions of $\s_6$.  The construction is identical to the one in the $n=6$ case.  We can consider any of the 24 equations in \eqref{n7 scattering eqs} and solve the corresponding linear system for the desired $\s_i$.  We arbitrarily choose the first 23 equations.  Then the desired $23\times 23$ determinant in $\s_3$ is constructed from the $24 \times 24$ determinant in \eqref{24x24 Sylvester} by removing the last column, multiplying the second row by $\s_3$ and adding that to the first row and finally by deleting the second row.  For the $\s_4$ polynomial we do the same but now instead of the second row we consider the third row and for the $\s_5$ polynomial we consider the seventh row according to the canonical order of our basis.

It is possible to expand \eqref{24x24 Sylvester} in terms of 4-brackets using its Laplace expansion.  We consider the first four columns of \eqref{24x24 Sylvester} and view this as a $24 \times 4$ matrix.  We then form all possible 4-brackets made out of any non-zero four rows of the aforementioned $24 \times 4$ matrix.  There are 70 possibilities.  We next form all possible non-zero 4-brackets made out of the columns 5,6,7,8 of \eqref{24x24 Sylvester}.  We repeat this until we exhaust all columns of \eqref{24x24 Sylvester}.  We end up with six groups of 70 4-brackets each.  We then form all the products that contain one 4-bracket of each group such that each row of any 4-bracket appears only once.  Of course in doing the Laplace expansion an appropriate sign needs to be considered.  We end up with an expression that contains 3274 terms of products of six 4-brackets each.  It is possible to express an expansion of \eqref{24x24 Sylvester} in terms of 4-brackets as a $6 \times 6$ determinant with elements 4-brackets.  This construction will be presented in the next section.

\subsection{$n=8$}
In order to easier understand the general case it is instructive to study the $n=8$ case that requires the construction of a $120 \times 120$ determinant.  We consider the map
\be 
\phi_5 ~ : ~ S(0,1,2,3)^5 \rightarrow S(1,2,3,4),~ (f_1,f_2,f_3,f_4,f_5) \mapsto \sum_{i=1}^5 f_i g_i.
\ee
The construction is similar to the previous cases.  We have worked it out and found that the 120 degree polynomial in $\s_7$ is given by the $120 \times 120$ determinant
\be \label{120x120 determinant}
\begin{vmatrix}
C_0 & 0 & 0 & 0 \\
C_1 & C_0 & 0 & 0 \\
0 & C_1 & C_0 & 0 \\
0 & 0 & C_1 & C_0 \\
0 & 0 & 0 & C_1
\end{vmatrix}
=0,
\ee 
where 
\be 
C_{j_1}=
\begin{pmatrix}
B_{0 j_1} & 0 & 0 \\
B_{1 j_1} & B_{0 j_1} & 0 \\
0 & B_{1 j_1} & B_{0 j_1} \\
0 & 0 & B_{1 j_1}
\end{pmatrix},
\qquad
B_{j_1 j_2}=
\begin{pmatrix}
A_{0j_1 j_2} & 0 \\
A_{1j_1 j_2} & A_{0j_1 j_2} \\
0 & A_{1j_1 j_2}
\end{pmatrix}
\ee
and
\be
A_{j_1 j_2 j_3} = 
\begin{pmatrix}
c_{0 j_1 j_2 j_3 1} & c_{0 j_1 j_2 j_3 2} & c_{0 j_1 j_2 j_3 3} & c_{0 j_1 j_2 j_3 4} & c_{0 j_1 j_2 j_3 5} \\
c_{1 j_1 j_2 j_3 1} & c_{1 j_1 j_2 j_3 2} & c_{1 j_1 j_2 j_3 3} & c_{1 j_1 j_2 j_3 4} & c_{1 j_1 j_2 j_3 5}
\end{pmatrix},\qquad j_i = 0,1.
\ee
An expansion in terms of 5-brackets can be constructed by considering groups of five columns of \eqref{120x120 determinant}, forming all possible 5-brackets and then forming products of the 5-brackets.  The final answer is a sum and difference of terms each one containing the product of 24 5-brackets.

\subsection{General case}
For general $n$ we consider the map
\be 
\phi_{n-3} ~ : ~ S(0,1,\ldots, n-5)^{n-3} \rightarrow S(1,2,\ldots, n-4 ),~ (f_1,f_2,\ldots, f_{n-3}) \mapsto \sum_{i=1}^{n-3} f_i g_i.
\ee
The dimension of the range and image of $\phi_{n-3}$ is $(n-3)!$.  We introduce the notation
\be \label{recursion 1}
A^{(2)}_{j_1 j_2 \ldots j_{n-5}} =
\begin{pmatrix}
c_{0 j_1 j_2 \ldots j_{n-5} 1} & c_{0 j_1 j_2 \ldots j_{n-5} 2} & \cdots & c_{0 j_1 j_2 \ldots j_{n-5} (n-3)} \\  
c_{1 j_1 j_2 \ldots j_{n-5} 1} & c_{1 j_1 j_2 \ldots j_{n-5} 2} & \cdots & c_{1 j_1 j_2 \ldots j_{n-5} (n-3)}
\end{pmatrix}
\ee
and
\be\label{recursion 2}
A^{(m)}_{j_1 j_2 \ldots j_{n-m-3}} = 
\underbracedmatrix{
A^{(m-1)}_{0 j_1 j_2 \ldots j_{n-m-3}} & 0 & \cdots & 0 \\[0.2cm]
A^{(m-1)}_{1 j_1 j_2 \ldots j_{n-m-3}} & A^{(m-1)}_{0 j_1 j_2 \ldots j_{n-m-3}} & \cdots & 0 \\[0.2cm]
0 & A^{(m-1)}_{1 j_1 j_2 \ldots j_{n-m-3}} & \cdots & 0 \\[0.2cm]
\vdots & \vdots & \ddots & \vdots \\[0.2cm]
0 & 0 & \cdots & A^{(m-1)}_{1 j_1 j_2 \ldots j_{n-m-3}} 
}{m-1}, m \geq 2.
\ee
Then the $(n-3)!$ order polynomial in $\s_{n-1}$ is given by the determinant
\be 
{\rm det} \, A^{(n-3)} = 0.
\ee
It is easy to solve the recursion \eqref{recursion 1}, \eqref{recursion 2} and explicitly write down $A^{(n-3)}$.  The final answer is a repeated structure isolated from the non-zero rows of the first $(n-3)$ columns.  The elimination of the rest of the variables as well as the expansion on $A^{(n-3)}$ in terms of $n$-brackets follows from the discussion of the $n=7$ case.  We should also comment that the determinant of $A^{(n-3)}$ can be associated to the Cayley hyperdeterminant of a hypermatrix of the boundary format with entries the coefficients of \eqref{generalized scattering equations 2}.

\subsection{How to evaluate the amplitudes}
One of the motivations of this work was to find a way to explicitly evaluate the scattering amplitudes using the scattering equations.  We have succeeded in doing something more.  In fact one can in principle explicitly evaluate any expression of the form
\be \label{sum of roots}
\sum_{\rm roots} f(\s_i),
\ee 
where the sum runs over the solutions of the generalized scattering equations \eqref{generalized scattering equations 1} and $f(\s_i)$ is an arbitrary rational function of the variables that appear in the scattering equations.  The idea is simple to understand and it was first presented in \cite{Kalousios:2015fya}, enhanced in \cite{Huang:2015yka} and further elaborated in \cite{Cardona:2015eba}.  One uses the elimination to express the function $f(\s_i)$ in terms of $\s_{n-1}$ only.  We call the resulted function $g(\s_{n-1})$.  Then, the sum of roots of \eqref{sum of roots} is given purely in terms of the coefficients of the $(n-3)!$ order polynomial in $\s_{n-1}$ using the Vieta formulae for the roots of a polynomial.  Furthermore, it is a rational function.  A way to efficiently do this is via the help of the companion matrices (\cite{Huang:2015yka} and \cite{Cardona:2015eba}).  One replaces  $\s_{n-1}$ in $g(\s_{n-1})$ by its companion matrix, $T(\s_{n-1})$, performs the multiplication and summation of the matrices and then takes the trace of the final matrix.  Schematically we have
\be \label{sum of roots 2}
\sum_{\rm roots} f(\s_i) = \sum_{\rm roots} g(\s_{n-1}) = {\rm Tr} \left[g(T(\s_{n-1}))\right].
\ee
For more details and for some examples we refer the reader to the aforementioned literature. 

\section{Elimination in the scattering equations via B\'{e}zout type determinants}

As we have seen in the previous section, the Sylvester determinant is an elegant way to compute the resultant of a set of multivariate polynomials.  In the multilinear case we considered we ended up with determinants of size $(n-3)!$.  The Sylvester method is by no means the only way to use elimination.  In this section we consider an alternative to Sylvester and we discuss the construction of the so-called B\'{e}zout formula for the scattering equations.  The resultant coming from the determinant of the B\'{e}zout formula has dimension $(n-4)!$ and it is equivalent to the one obtained in the previous section.  The elements of the B\'{e}zout formula are immediately given in terms of $(n-3)$-brackets.  Other mixed approaches of intermediate dimensionality can also be considered but we will not discuss them here.

Although the construction of the B\'ezout formula applies straightforwardly to general $n$， with a factorially increasing difficulty, for the sake of simplicity we are going to explicitly consider only the particular cases $n=6$ and $n=7$.  The general idea is simple and was originally considered in \cite{Dickenstein} although in this section we will follow \cite{Sturmfels:2002}. 

For a polynomial set of the general form (\ref{generalized scattering equations 2}) we define the matrix whose entries are given by
\be\ba\label{bezoutian}
M_{1m}&=g_m(\s_3,\ldots,\s_{n-2}),\quad m=1,2,\ldots,n-3,\,\,\,i=2,\ldots,n-3,\\
M_{im}&=\frac{g_m(\tilde{\s}_3,\ldots,\tilde{\s}_{i+1},\s_{i+2},\ldots,\s_{n-2})-g_m(\tilde{\s}_3,\ldots,\tilde{\s}_{i},\s_{i+1},\ldots,\s_{n-2})}{\tilde{\s}_i-\sigma_i}\,.
\ea\ee
The determinant of this matrix is a polynomial known as the affine b\'ezoutian, which we denote as $B(\s,\tilde{\s})$. It can be proved that the monomials 
\be\label{monomials}\s_3^{\alpha_3}\cdots\s_{n-2}^{\alpha_{n-2}}\tilde{\s}_3^{\beta_3}\cdots\tilde{\s}_{n-2}^{\beta_{n-2}}\,,\ee 
composing the affine b\'ezoutian satisfy the following relation,
\be\label{order relations Bezout} \alpha_i<i-2,\quad \beta_i<(n-2)-i\,,\ee
and hence,
\be B\in S(0,1,\ldots,n-5)\otimes S(n-5,n-6,\ldots,0)^*\,.\ee
This in turn implies that the affine b\'ezoutian is independent of $\sigma_3$ and $\tilde{\s}_{n-2}$. In the formulae that follow we subtract $\s_3$ times the second row from the first row in \eqref{bezoutian} in order to explicitly eliminate $\s_3$.  This is possible because ${\rm det}\,M=0$ upon use of the scattering equations.  More importantly, the affine b\'ezoutian  can be interpreted as a linear map from the dual vector space $S(n-5,n-6,\ldots,0)^*$ to $S(0,1,\ldots,n-5)$, which is represented by a $(n-4)!\times (n-4)!$ matrix $B$, i.e, the coefficients of the affine b\'ezoutian are the components of the aforementioned matrix $B$, which in the mathematical literature is called the B\'ezout formula.  
Since the components of the B\'ezout formula come from the computation of a determinant, namely ${\rm det}\,M$, they can be written in terms of brackets. Here we sketch one way to obtain them.  

First choose an order for the monomial basis and think of it as a vector. Take into account that there are two such vectors, one in the basis built out of $\sigma$s for $S(0,1,\ldots,n-5)$ and the other in the dual basis built out of $\tilde{\sigma}$s for the dual space $S(n-5,n-6,\ldots,0)^*$. 

Let us consider the bracket composition of the $(l,m)^{\rm th}$ element of the matrix $B$, namely $B_{lm}$. The 
$l^{\rm th}\otimes m^{\rm th}$ component of the basis $S(0,1,\ldots,n-5)\otimes S(n-5,n-6,\ldots,0)^*$ is given by an element in the set (\ref{monomials}) with values for $\alpha_i$ and $\beta_j$ that depend on the chosen order or our monomial basis.  Then we have
\be
 B_{lm}=\left[\prod^{n-2}_{i=4}\frac{1}{\a_i!}\left(\partial_{\s_i}\right)^{\alpha_i}\prod^{n-3}_{j=3}\frac{1}{\b_j!}\left(\partial_{\tilde{\s}_j}\right)^{\beta_j}\right]{\rm det}(M)|_{(\s,\tilde{\s})=0}\,.
\ee
 The notation $|_{(\s,\tilde{\s})=0}$ means we set all the $\s$s and $\tilde{\s}$s to zero after taking the derivatives.
Let us now move to some particular examples.

\subsection{$n=6$}
In the case of six particles, the polynomial form of the scattering equations is given by \eqref{n6 scattering eqs}.  For this polynomial set, the $i^{\rm th}$ column of the $M$ matrix is given by,
\be\label{6x6 M determinant}
M_6=\left(\begin{matrix}
c_{00i} + c_{01i}\s_4  \\
c_{10i}+c_{11i} \s_4\\
c_{01i}+c_{11i} \tilde{\s}_3 \\
\end{matrix} \right)\,.
\ee
The determinant of $M$ is given by a polynomial in $S(0,1)\otimes S(1,0)^*$ and the linear map from $S(1,0)^*$ to $S(0,1)$ in the basis $\{1,\s_4\}\otimes\{1,\tilde{\s}_3\}$ can be conveniently written as ,
\be\label{6x6 B determinant}
B_6=\left(\begin{matrix}
[00,10,01] & [00,10,11] \\
[00,01,11] & [10,01,11] \\
\end{matrix}\right)\,,
\ee 
where the 3-brackets were defined in \eqref{n-bracket}.  The determinant of $B_6$ produces the same resultant as the one given in (\ref{6x6 determinant}).

\subsection{$n=7$}
The scattering equations are given by \eqref{n7 scattering eqs}.
For this polynomial set, the $i^{\rm th}$ column of the $M$ matrix is given by,
\be\label{24x24 M determinant}
M_7=\left(\begin{matrix}
c_{000i} + c_{001i}\s_5+c_{010i}\s_4+c_{011i}\s_4\s_5  \\
c_{100i} + c_{101i}\s_5+c_{110i}\s_4+c_{111i}\s_4\s_5  \\
c_{010i} + c_{011i}\s_5+c_{110i}\tilde{\s}_3+c_{011i}\tilde{\s}_3\s_5  \\
c_{001i} + c_{011i}\tilde{\s}_4+c_{101i}\tilde{\s}_3+c_{111i}\tilde{\s}_3\tilde{\s}_4  \\
\end{matrix} \right)\,.
\ee
We compute the matrix of the linear map from $S(2,1,0)^*$ to $S(0,1,2)$ in the basis $(1,\s_5, \s_5^2, \s_4, \s_4\s_5, \s_4 \s_5^2)\otimes(1,\tilde{\s}_4,\tilde{\s}_3,\tilde{\s}_4\tilde{\s}_3, \tilde{\s}_3^2,\tilde{\s}_4\tilde{\s}_3^2)$,  to be
\be 
B_7=
\begin{pmatrix}
A_{4 \times 4} & B_{4 \times 2} \\
C_{2 \times 4} & D_{2 \times 2}
\end{pmatrix}
\ee
with
\ben \small{
A=
\begin{pmatrix}
[000,001,010,100]	& [000,010,011,100]	&[000,100,101,110]	&[000,100,110,111]\\[0.2cm]
-[000,001,011,101]	&[001,010,011,101]	&[001,100,101,111]	&[001,101,110,111]\\[0.2cm]
-[000,001,010,110]	&[000,010,011,110]	&[010,100,101,110]	&[010,100,110,111]\\[0.2cm]
-[000,001,011,111]	&[001,010,011,111] &[011,100,101,111]	&[011,101,110,111]
\end{pmatrix}},
\een
\ben \small{
B=
\begin{pmatrix}
[000,001,100,110]-[000,010,100,101]&	[000,011,100,110]-[000,010,100,111]\\[0.2cm]
[000,001,101,111]-[001,011,100,101]&	[001,011,101,110]-[001,010,101,111]\\[0.2cm]
[000,010,101,110]-[001,010,100,110]&	[010,011,100,110]-[000,010,110,111]\\[0.2cm]
[000,011,101,111]-[001,011,100,111]&	[010,011,101,111]-[001,011,110,111]
\end{pmatrix}},
\een
\ben \small{
C=
\begin{pmatrix}
-[000,001,010,101]	& [000,010,011,101]&	[000,100,101,111]&	[000,101,110,111]\\
-[000,001,011,100]	& +[001,010,011,100]&	+[001,100,101,110]&	+[001,100,110,111]\\[0.3cm]
-[000,001,010,111]&	[000,010,011,111]&	[010,100,101,111]&	[010,101,110,111]\\
-[000,001,011,110]&	+[001,010,011,110]&	+[011,100,101,110]&	+[011,100,110,111]
\end{pmatrix}},
\een
\ben \small{
D=
\begin{pmatrix}
[000,001,100,111]+[000,001,101,110]&	-[000,010,101,111]+[000,011,101,110]\\
-[000,011,100,101]-[001,010,100,101]&	-[001,010,100,111]+[001,011,100,110]\\[0.3cm]
[000,010,101,111]+[000,011,101,110]&	-[000,011,110,111]-[001,010,110,111]\\
-[001,010,100,111]-[001,011,100,110]&	+[010,011,100,111]+[010,011,101,110]
\end{pmatrix}},
\een
where we have used the bracket notation \eqref{n-bracket}.

The expansion of the determinant above gives us the 24 degree polynomial in $\s_6$ corresponding to the resultant of the system (\ref{n7 scattering eqs}).

\section{Discussion}
In this work we applied the classical elimination theory in the context of the scattering equations.  We chose to use a generalized set of the scattering equations in their polynomial form which leads to more compact expressions.  We have achieved to construct the one variable polynomial of degree $(n-3)!$ in one of the variables of the scattering equations.  The answer is given compactly by the determinant of a $(n-3)! \times (n-3)!$ matrix of Sylvester type.  Then we expressed the rest of the variables of the scattering equations as function of one of the variables.  The determinant we found satisfies on-shell recursion relations and it would be interesting to study its relation to BCFW recursions \cite{Britto:2004ap,Britto:2005fq}.

One of the features of our relations is that one can interestingly expand our expressions in terms of $n$-brackets that can be viewed as Pl\"{u}cker coordinates on the variety of lines in $P^n$.  One way to see this is to use a particular Laplace expansion of the Sylvester determinant.  Then the Sylvester determinant has the interpretation of the Chow form of the subvariety $P^1 \times P^1 \times \cdots \times P^1$ in the Segre embedding \cite{Gelfand}.

There is an alternative way to use the elimination, namely the B\'{e}zout construction.  In this case one ends up with a $(n-4)! \times (n-4)!$ determinant with elements constructed out of $(n-3)$-brackets.  The expansion of the B\'{e}zout determinant gives the same polynomials as the Sylvester determinant after making use of Pl\"{u}cker relations.  Mixed determinantal expressions of intermediate dimensionality should also exist but we did not investigate this here.

Our derivation plays the role of the construction of a Gr\"{o}bner basis of the scattering equations.  One of the applications of our results is the evaluation of the sum over the solutions of the generalized scattering equations of any rational expression of the variables that appear in the scattering equation.  This can now be achieved by pure algebraic manipulations.  The idea is based on the fact that the sum over roots is given by particular combinations of the coefficients of the scattering equations via simple algebra.  Therefore, one does not need to explicitly know the solutions of the equations.  In fact one cannot know explicitly the solutions in view of the Abel-Ruffini theorem.  In the case of the scattering amplitudes, we hope that specialized versions of the elimination could lead to huge calculational simplifications.

There is further a connection of the Sylvester determinant and the Cayley hyperdeterminant of a particular hypermatrix with elements the coefficients of the scattering equations.

One could hope that our construction can lead to a further investigation of the scattering equations and amplitudes and even perhaps connections with different geometrical and combinatorial structures.

\vspace{5mm}

\noindent
{\bf Acknowledgments}

\vspace{3mm}

\noindent
It is a pleasure to thank Humberto Gomez and Francisco Rojas for useful comments and discussions.  The work of C.C. is supported in part by the National Center for Theoretical Sciences (NCTS), Taiwan.  The work of C.K. is supported by the S\~ao Paulo Research Foundation (FAPESP) under grants 2011/11973-4 and 2012/00756-5.

\bibliographystyle{utphys}
\bibliography{mybib}

\providecommand{\href}[2]{#2}\begingroup\raggedright\begin{thebibliography}{10}

\bibitem{Witten:2003nn}
E.~Witten, ``\emph{Perturbative gauge theory as a string theory in twistor
  space},'' \href{http://dx.doi.org/10.1007/s00220-004-1187-3}{{\em Commun.
  Math. Phys.} {\bfseries 252} (2004) 189--258},
\href{http://arxiv.org/abs/hep-th/0312171}{{\ttfamily arXiv:hep-th/0312171
  [hep-th]}}.

\bibitem{Cachazo:2013gna}
F.~Cachazo, S.~He, and E.~Y. Yuan, ``\emph{Scattering equations and
  Kawai-Lewellen-Tye orthogonality},''
  \href{http://dx.doi.org/10.1103/PhysRevD.90.065001}{{\em Phys.Rev.}
  {\bfseries D90} no.~6, (2014) 065001},
\href{http://arxiv.org/abs/1306.6575}{{\ttfamily arXiv:1306.6575 [hep-th]}}.

\bibitem{Cachazo:2013hca}
F.~Cachazo, S.~He, and E.~Y. Yuan, ``\emph{Scattering of Massless Particles in
  Arbitrary Dimensions},''
  \href{http://dx.doi.org/10.1103/PhysRevLett.113.171601}{{\em Phys.Rev.Lett.}
  {\bfseries 113} no.~17, (2014) 171601},
\href{http://arxiv.org/abs/1307.2199}{{\ttfamily arXiv:1307.2199 [hep-th]}}.

\bibitem{Cachazo:2013iea}
F.~Cachazo, S.~He, and E.~Y. Yuan, ``\emph{Scattering of Massless Particles:
  Scalars, Gluons and Gravitons},''
  \href{http://dx.doi.org/10.1007/JHEP07(2014)033}{{\em JHEP} {\bfseries 1407}
  (2014) 033},
\href{http://arxiv.org/abs/1309.0885}{{\ttfamily arXiv:1309.0885 [hep-th]}}.

\bibitem{Cachazo:2014nsa}
F.~Cachazo, S.~He, and E.~Y. Yuan, ``\emph{Einstein-Yang-Mills Scattering
  Amplitudes From Scattering Equations},''
  \href{http://dx.doi.org/10.1007/JHEP01(2015)121}{{\em JHEP} {\bfseries 1501}
  (2015) 121},
\href{http://arxiv.org/abs/1409.8256}{{\ttfamily arXiv:1409.8256 [hep-th]}}.

\bibitem{Cachazo:2014xea}
F.~Cachazo, S.~He, and E.~Y. Yuan, ``\emph{Scattering Equations and Matrices:
  From Einstein To Yang-Mills, DBI and NLSM},''
\href{http://arxiv.org/abs/1412.3479}{{\ttfamily arXiv:1412.3479 [hep-th]}}.

\bibitem{Dolan:2013isa}
L.~Dolan and P.~Goddard, ``\emph{Proof of the Formula of Cachazo, He and Yuan
  for Yang-Mills Tree Amplitudes in Arbitrary Dimension},''
  \href{http://dx.doi.org/10.1007/JHEP05(2014)010}{{\em JHEP} {\bfseries 1405}
  (2014) 010},
\href{http://arxiv.org/abs/1311.5200}{{\ttfamily arXiv:1311.5200 [hep-th]}}.

\bibitem{Fairlie}
D.~Fairlie and D.~Roberts, {\em Dual models without tachyons - a new approach,
  \rm{(unpublished Durham preprint PRINT-72-2440, 1972)}}.

\bibitem{Roberts}
D.~Roberts, {\em Mathematical Structure of Dual Amplitudes, \rm{(Durham PhD
  thesis, 1972) p.73 f. [available at Durham E-Theses online]}}.

\bibitem{Fairlie:2008dg}
D.~B. Fairlie, ``\emph{A Coding of Real Null Four-Momenta into World-Sheet
  Co-ordinates},'' \href{http://dx.doi.org/10.1155/2009/284689}{{\em
  Adv.Math.Phys.} {\bfseries 2009} (2009) 284689},
\href{http://arxiv.org/abs/0805.2263}{{\ttfamily arXiv:0805.2263 [hep-th]}}.

\bibitem{Gross:1987ar}
D.~J. Gross and P.~F. Mende, ``\emph{String Theory Beyond the Planck Scale},''
\href{http://dx.doi.org/10.1016/0550-3213(88)90390-2}{{\em Nucl.Phys.}
  {\bfseries B303} (1988) 407}.

\bibitem{Witten:2004cp}
E.~Witten, ``\emph{Parity invariance for strings in twistor space},''
  \href{http://dx.doi.org/10.4310/ATMP.2004.v8.n5.a1}{{\em
  Adv.Theor.Math.Phys.} {\bfseries 8} (2004) 779--796},
\href{http://arxiv.org/abs/hep-th/0403199}{{\ttfamily arXiv:hep-th/0403199
  [hep-th]}}.

\bibitem{Caputa:2011zk}
P.~Caputa and S.~Hirano, ``\emph{Observations on Open and Closed String
  Scattering Amplitudes at High Energies},''
  \href{http://dx.doi.org/10.1007/JHEP02(2012)111}{{\em JHEP} {\bfseries 1202}
  (2012) 111},
\href{http://arxiv.org/abs/1108.2381}{{\ttfamily arXiv:1108.2381 [hep-th]}}.

\bibitem{Caputa:2012pi}
P.~Caputa, ``\emph{Lightlike contours with fermions},''
  \href{http://dx.doi.org/10.1016/j.physletb.2012.09.006}{{\em Phys.Lett.}
  {\bfseries B716} (2012) 475--480},
\href{http://arxiv.org/abs/1205.6369}{{\ttfamily arXiv:1205.6369 [hep-th]}}.

\bibitem{Makeenko:2011dm}
Y.~Makeenko and P.~Olesen, ``\emph{The QCD scattering amplitude from area
  behaved Wilson loops},''
  \href{http://dx.doi.org/10.1016/j.physletb.2012.02.032}{{\em Phys.Lett.}
  {\bfseries B709} (2012) 285--288},
\href{http://arxiv.org/abs/1111.5606}{{\ttfamily arXiv:1111.5606 [hep-th]}}.

\bibitem{Cachazo:2012uq}
F.~Cachazo, ``\emph{Fundamental BCJ Relation in $N=4$ SYM From The Connected
  Formulation},''
\href{http://arxiv.org/abs/1206.5970}{{\ttfamily arXiv:1206.5970 [hep-th]}}.

\bibitem{Dolan:2014ega}
L.~Dolan and P.~Goddard, ``\emph{The Polynomial Form of the Scattering
  Equations},'' \href{http://dx.doi.org/10.1007/JHEP07(2014)029}{{\em JHEP}
  {\bfseries 1407} (2014) 029},
\href{http://arxiv.org/abs/1402.7374}{{\ttfamily arXiv:1402.7374 [hep-th]}}.

\bibitem{Naculich:2014naa}
S.~G. Naculich, ``\emph{Scattering equations and BCJ relations for gauge and
  gravitational amplitudes with massive scalar particles},''
  \href{http://dx.doi.org/10.1007/JHEP09(2014)029}{{\em JHEP} {\bfseries 1409}
  (2014) 029},
\href{http://arxiv.org/abs/1407.7836}{{\ttfamily arXiv:1407.7836 [hep-th]}}.

\bibitem{Weinzierl:2014vwa}
S.~Weinzierl, ``\emph{On the solutions of the scattering equations},''
  \href{http://dx.doi.org/10.1007/JHEP04(2014)092}{{\em JHEP} {\bfseries 1404}
  (2014) 092},
\href{http://arxiv.org/abs/1402.2516}{{\ttfamily arXiv:1402.2516 [hep-th]}}.

\bibitem{Kalousios:2013eca}
C.~Kalousios, ``\emph{Massless scattering at special kinematics as Jacobi
  polynomials},'' \href{http://dx.doi.org/10.1088/1751-8113/47/21/215402}{{\em
  J.Phys.} {\bfseries A47} (2014) 215402},
\href{http://arxiv.org/abs/1312.7743}{{\ttfamily arXiv:1312.7743 [hep-th]}}.

\bibitem{Lam:2014tga}
C.~Lam, ``\emph{Permutation Symmetry of the Scattering Equations},''
  \href{http://dx.doi.org/10.1103/PhysRevD.91.045019}{{\em Phys.Rev.}
  {\bfseries D91} no.~4, (2015) 045019},
\href{http://arxiv.org/abs/1410.8184}{{\ttfamily arXiv:1410.8184 [hep-th]}}.

\bibitem{He:2014wua}
Y.-H. He, C.~Matti, and C.~Sun, ``\emph{The Scattering Variety},''
  \href{http://dx.doi.org/10.1007/JHEP10(2014)135}{{\em JHEP} {\bfseries 10}
  (2014) 135},
\href{http://arxiv.org/abs/1403.6833}{{\ttfamily arXiv:1403.6833 [hep-th]}}.

\bibitem{Schwab:2014xua}
B.~U.~W. Schwab and A.~Volovich, ``\emph{Subleading Soft Theorem in Arbitrary
  Dimensions from Scattering Equations},''
  \href{http://dx.doi.org/10.1103/PhysRevLett.113.101601}{{\em Phys. Rev.
  Lett.} {\bfseries 113} no.~10, (2014) 101601},
\href{http://arxiv.org/abs/1404.7749}{{\ttfamily arXiv:1404.7749 [hep-th]}}.

\bibitem{Cachazo:2015ksa}
F.~Cachazo, S.~He, and E.~Y. Yuan, ``\emph{New Double Soft Emission
  Theorems},'' \href{http://dx.doi.org/10.1103/PhysRevD.92.065030}{{\em Phys.
  Rev.} {\bfseries D92} no.~6, (2015) 065030},
\href{http://arxiv.org/abs/1503.04816}{{\ttfamily arXiv:1503.04816 [hep-th]}}.

\bibitem{Afkhami-Jeddi:2014fia}
N.~Afkhami-Jeddi, ``\emph{Soft Graviton Theorem in Arbitrary Dimensions},''
\href{http://arxiv.org/abs/1405.3533}{{\ttfamily arXiv:1405.3533 [hep-th]}}.

\bibitem{Kalousios:2014uva}
C.~Kalousios and F.~Rojas, ``\emph{Next to subleading soft-graviton theorem in
  arbitrary dimensions},''
  \href{http://dx.doi.org/10.1007/JHEP01(2015)107}{{\em JHEP} {\bfseries 01}
  (2015) 107},
\href{http://arxiv.org/abs/1407.5982}{{\ttfamily arXiv:1407.5982 [hep-th]}}.

\bibitem{Naculich:2015zha}
S.~G. Naculich, ``\emph{CHY representations for gauge theory and gravity
  amplitudes with up to three massive particles},''
\href{http://arxiv.org/abs/1501.03500}{{\ttfamily arXiv:1501.03500 [hep-th]}}.

\bibitem{Naculich:2015coa}
S.~G. Naculich, ``\emph{Amplitudes for massive vector and scalar bosons in
  spontaneously-broken gauge theory from the CHY representation},''
  \href{http://dx.doi.org/10.1007/JHEP09(2015)122}{{\em JHEP} {\bfseries 09}
  (2015) 122},
\href{http://arxiv.org/abs/1506.06134}{{\ttfamily arXiv:1506.06134 [hep-th]}}.

\bibitem{Weinzierl:2014ava}
S.~Weinzierl, ``\emph{Fermions and the scattering equations},''
\href{http://arxiv.org/abs/1412.5993}{{\ttfamily arXiv:1412.5993 [hep-th]}}.

\bibitem{delaCruz:2015raa}
L.~de~la Cruz, A.~Kniss, and S.~Weinzierl, ``\emph{The CHY representation of
  tree-level primitive QCD amplitudes},''
\href{http://arxiv.org/abs/1508.06557}{{\ttfamily arXiv:1508.06557 [hep-th]}}.

\bibitem{Mason:2013sva}
L.~Mason and D.~Skinner, ``\emph{Ambitwistor strings and the scattering
  equations},'' \href{http://dx.doi.org/10.1007/JHEP07(2014)048}{{\em JHEP}
  {\bfseries 1407} (2014) 048},
\href{http://arxiv.org/abs/1311.2564}{{\ttfamily arXiv:1311.2564 [hep-th]}}.

\bibitem{Berkovits:2013xba}
N.~Berkovits, ``\emph{Infinite Tension Limit of the Pure Spinor Superstring},''
  \href{http://dx.doi.org/10.1007/JHEP03(2014)017}{{\em JHEP} {\bfseries 1403}
  (2014) 017},
\href{http://arxiv.org/abs/1311.4156}{{\ttfamily arXiv:1311.4156 [hep-th]}}.

\bibitem{Bjerrum-Bohr:2014qwa}
N.~E.~J. Bjerrum-Bohr, P.~H. Damgaard, P.~Tourkine, and P.~Vanhove,
  ``\emph{Scattering Equations and String Theory Amplitudes},''
  \href{http://dx.doi.org/10.1103/PhysRevD.90.106002}{{\em Phys. Rev.}
  {\bfseries D90} no.~10, (2014) 106002},
\href{http://arxiv.org/abs/1403.4553}{{\ttfamily arXiv:1403.4553 [hep-th]}}.

\bibitem{Casali:2015vta}
E.~Casali, Y.~Geyer, L.~Mason, R.~Monteiro, and K.~A. Roehrig, ``\emph{New
  Ambitwistor String Theories},''
  \href{http://dx.doi.org/10.1007/JHEP11(2015)038}{{\em JHEP} {\bfseries 11}
  (2015) 038},
\href{http://arxiv.org/abs/1506.08771}{{\ttfamily arXiv:1506.08771 [hep-th]}}.

\bibitem{Kalousios:2015fya}
C.~Kalousios, ``\emph{Scattering equations, generating functions and all
  massless five point tree amplitudes},''
  \href{http://dx.doi.org/10.1007/JHEP05(2015)054}{{\em JHEP} {\bfseries 05}
  (2015) 054},
\href{http://arxiv.org/abs/1502.07711}{{\ttfamily arXiv:1502.07711 [hep-th]}}.

\bibitem{Huang:2015yka}
R.~Huang, J.~Rao, B.~Feng, and Y.-H. He, ``\emph{An Algebraic Approach to the
  Scattering Equations},''
\href{http://arxiv.org/abs/1509.04483}{{\ttfamily arXiv:1509.04483 [hep-th]}}.

\bibitem{Cardona:2015eba}
C.~Cardona and C.~Kalousios, ``\emph{Comments on the evaluation of massless
  scattering},''
\href{http://arxiv.org/abs/1509.08908}{{\ttfamily arXiv:1509.08908 [hep-th]}}.

\bibitem{Sogaard:2015dba}
M.~Sogaard and Y.~Zhang, ``\emph{Scattering Equations and Global Duality of
  Residues},''
\href{http://arxiv.org/abs/1509.08897}{{\ttfamily arXiv:1509.08897 [hep-th]}}.

\bibitem{Cachazo:2015nwa}
F.~Cachazo and H.~Gomez, ``\emph{Computation of Contour Integrals on ${\cal
  M}_{0,n}$},''
\href{http://arxiv.org/abs/1505.03571}{{\ttfamily arXiv:1505.03571 [hep-th]}}.

\bibitem{Baadsgaard:2015voa}
C.~Baadsgaard, N.~E.~J. Bjerrum-Bohr, J.~L. Bourjaily, and P.~H. Damgaard,
  ``\emph{Integration Rules for Scattering Equations},''
  \href{http://dx.doi.org/10.1007/JHEP09(2015)129}{{\em JHEP} {\bfseries 09}
  (2015) 129},
\href{http://arxiv.org/abs/1506.06137}{{\ttfamily arXiv:1506.06137 [hep-th]}}.

\bibitem{Baadsgaard:2015ifa}
C.~Baadsgaard, N.~E.~J. Bjerrum-Bohr, J.~L. Bourjaily, and P.~H. Damgaard,
  ``\emph{Scattering Equations and Feynman Diagrams},''
  \href{http://dx.doi.org/10.1007/JHEP09(2015)136}{{\em JHEP} {\bfseries 09}
  (2015) 136},
\href{http://arxiv.org/abs/1507.00997}{{\ttfamily arXiv:1507.00997 [hep-th]}}.

\bibitem{Dickenstein}
A.~Dickenstein and I.~Emiris, ``\emph{Multihomogeneous resultant matrices},''
  {\em \rm{ISSAC 2002}} (2002) .
  \href{http://mate.dm.uba.ar/~alidick}{http://mate.dm.uba.ar/{\raise.17ex\hbox{$\scriptstyle\sim$}}alidick}.

\bibitem{Sturmfels:2002}
B.~Sturmfels, ``\emph{Solving Systems of Polynomial Equations},'' {\em \rm{Cbms
  Regional Conference Series in Mathematics. American Mathematical Society}}
  (2002) .
  \href{https://math.berkeley.edu/~bernd/cbms.pdf}{https://math.berkeley.edu/{\raise.17ex\hbox{$\scriptstyle\sim$}}bernd/cbms.pdf}.

\bibitem{Britto:2004ap}
R.~Britto, F.~Cachazo, and B.~Feng, ``{New recursion relations for tree
  amplitudes of gluons},''
  \href{http://dx.doi.org/10.1016/j.nuclphysb.2005.02.030}{{\em Nucl. Phys.}
  {\bfseries B715} (2005) 499--522},
\href{http://arxiv.org/abs/hep-th/0412308}{{\ttfamily arXiv:hep-th/0412308
  [hep-th]}}.

\bibitem{Britto:2005fq}
R.~Britto, F.~Cachazo, B.~Feng, and E.~Witten, ``\emph{Direct proof of
  tree-level recursion relation in Yang-Mills theory},''
  \href{http://dx.doi.org/10.1103/PhysRevLett.94.181602}{{\em Phys.Rev.Lett.}
  {\bfseries 94} (2005) 181602},
\href{http://arxiv.org/abs/hep-th/0501052}{{\ttfamily arXiv:hep-th/0501052
  [hep-th]}}.

\bibitem{Gelfand}
{I.M. Gelfand, M.M. Kapranov, A.V. Zelevinsky}, ``\emph{Discriminants,
  Resultants, and Multidimensional Determinants},'' {\em \rm{(Birkh\"{a}user,
  Boston, 1994)}} .

\end{thebibliography}\endgroup
\end{document}